\documentclass[preprint]{elsarticle}
\usepackage{amsmath}
\usepackage{amssymb}
\usepackage[final]{epsfig}
\usepackage{color}
\usepackage{graphicx}
\usepackage{lineno}
\newcommand{\beq}{\begin{equation}}
\newcommand{\eeq}{\end{equation}}
\newcommand{\bdi}{\begin{displaymath}}
\newcommand{\edi}{\end{displaymath}}
\newcommand{\no}{\nonumber}
\newcommand{\bea}{\begin{eqnarray}}
\newcommand{\eea}{\end{eqnarray}}

\newcommand{\ov}{\overline}

\newcommand{\om}{\omega}

\begin{document}


\begin{frontmatter}

\title{Time resolution limits in silicon sensors from Landau fluctuations and electronics noise}

\author[]{W. Riegler}

\address{CERN}

\begin{abstract}
 
 \noindent 
In this report, we derive analytical expressions for the time resolution limits of standard silicon sensors, LGADs, and 3D trench sensors. We separately examine the effects of Landau fluctuations and electronic noise. To analyze Landau fluctuations, we relate the time resolution of a single electron-hole pair generated at a random position in the sensor to the time resolution associated with the full ionization pattern produced by a charged particle. For electronic noise, we explore optimal filtering techniques that minimize its impact on time resolution, and evaluate how closely these can be approximated by practical  filters. Finally, we demonstrate that the combined effect of Landau fluctuations and electronic noise cannot, in general, be simply expressed as the quadratic sum of the individual contributions.

\end{abstract}

\end{frontmatter}

\section{Introduction}

\noindent
In \cite{werner1} the fluctuation of the centroid time of the sensor signal was identified as a key quantity defining the intrinsic time resolution of a silicon sensor.  It was shown that this picture applies if the preamplifier integration time is of the same order or longer than the signal duration. Since the fluctuations arise from charge deposit fluctuations called 'Landau fluctuations', this contribution is sometimes called 'Landau noise'. 
The expressions in \cite{werner1} were derived for situations with a large number of e-h pairs, where this number can be assumed to be a continuous variable instead of a discrete one. In this report the exact expressions based on the discrete nature of the charge deposit are derived and they are therefore valid for arbitrarily thin sensors. It is also  shown that one only has to calculate the time resolution for a single e-h pair with a uniformly distributed random position in the sensor, which directly yields the time resolution for a charged particle passing the sensor. 
We will first review the characteristics of charge deposit in silicon and then calculate the fluctuations of the centroid time for different detector geometries. We then discuss the effect of Landau fluctuations for arbitrary amplifier bandwidth and thresholds. For large numbers of primary charges  the results derived in this report are equal to the ones derived in \cite{werner1}. However they are also valid for arbitrarily small number of primary charge deposits and therefore applicable to a wider range of sensors. \\ \\
To study the impact of electronic noise on time resolution, commonly referred to as {\it time jitter} or simply {\it jitter}, we begin by analyzing the average silicon signal with superimposed series noise and evaluate optimal filters that maximize the signal slope-to-noise ratio. We then examine how closely these optimal filters can be approximated by practical amplifier designs.


\section{Charge Deposit in Silicon}

\begin{figure}[ht]
 \begin{center}
     a)\includegraphics[width=4.5cm]{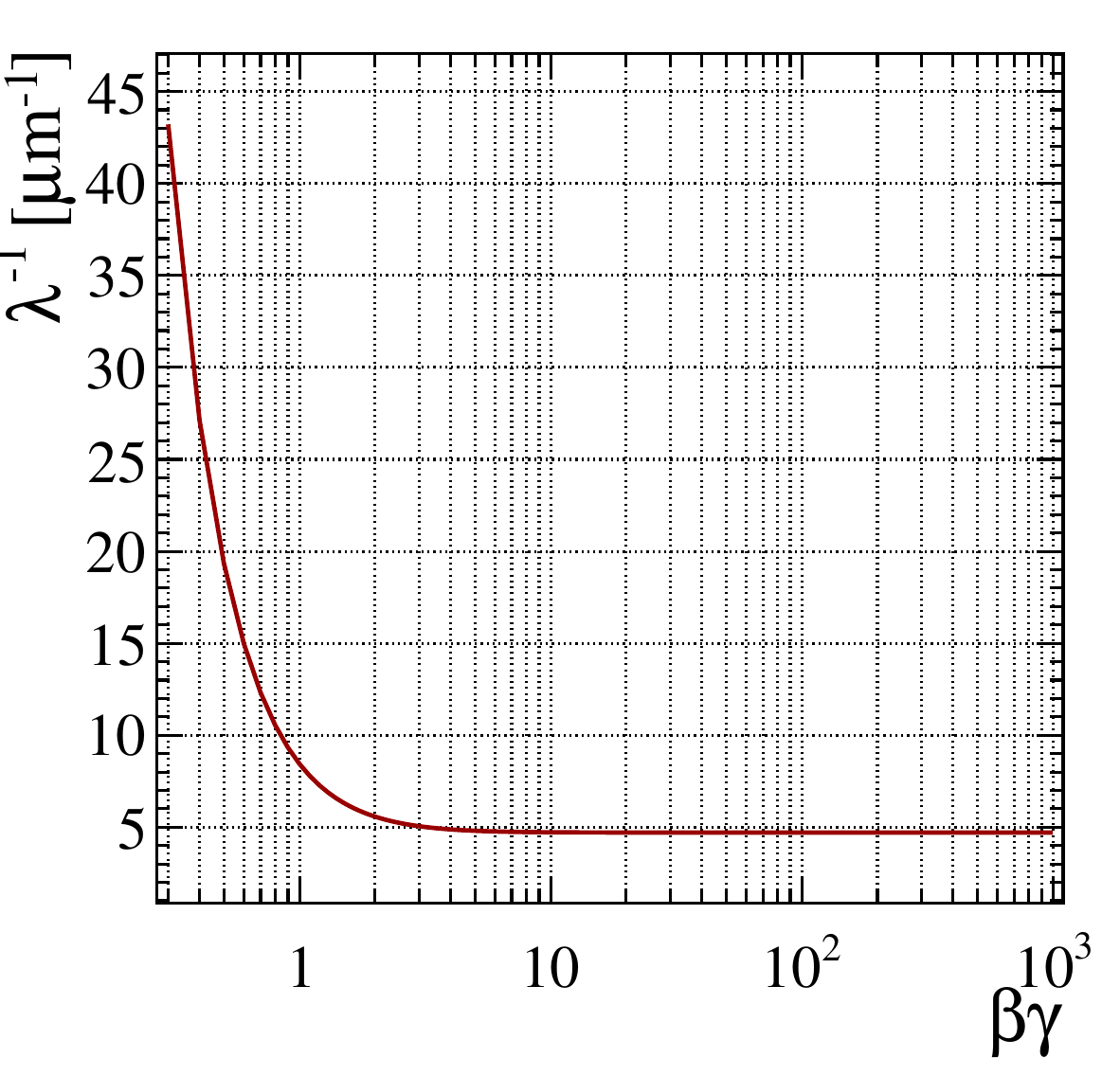}
     b)\includegraphics[width=4.5cm]{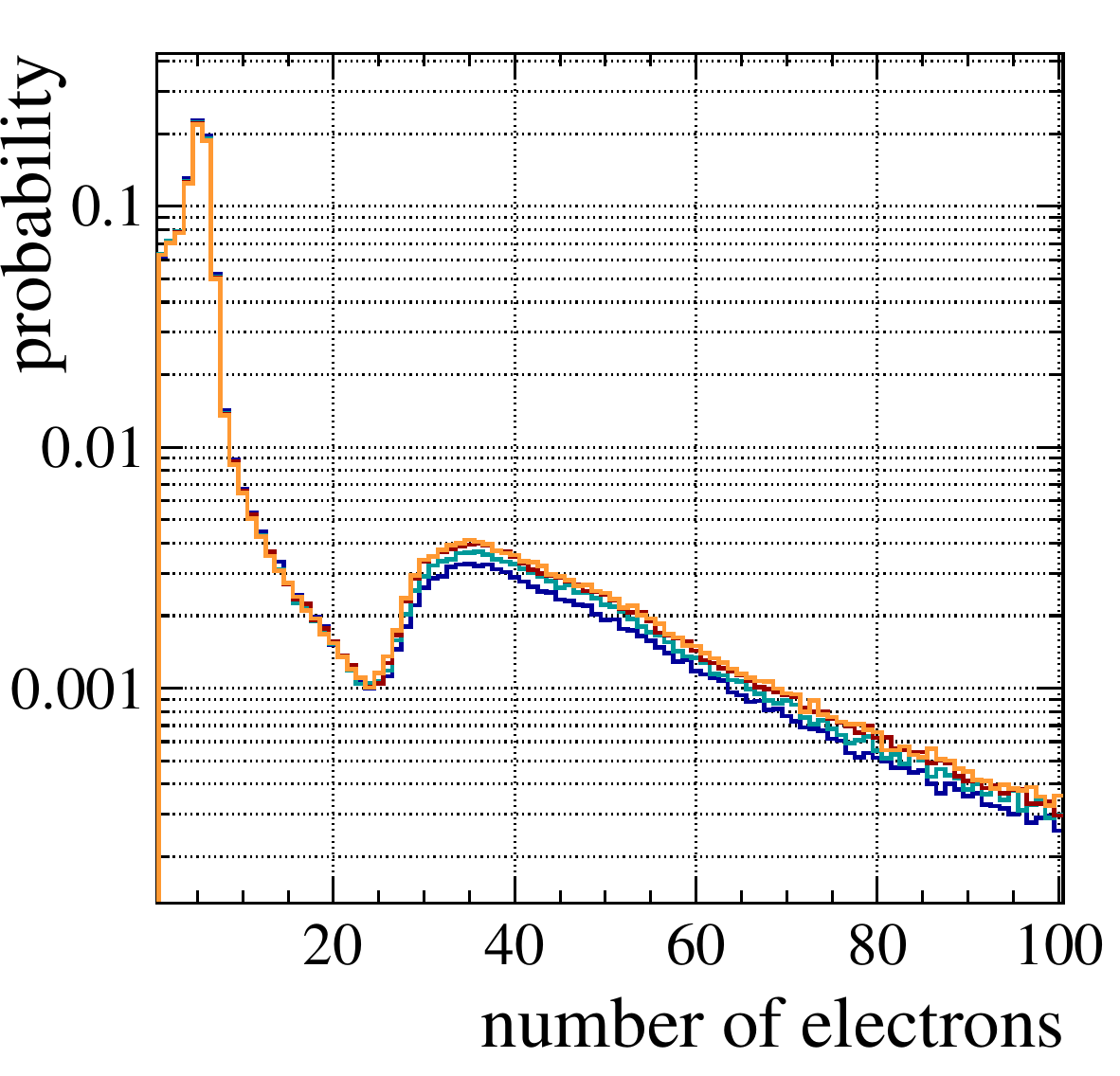}
      c)\includegraphics[width=4.5cm]{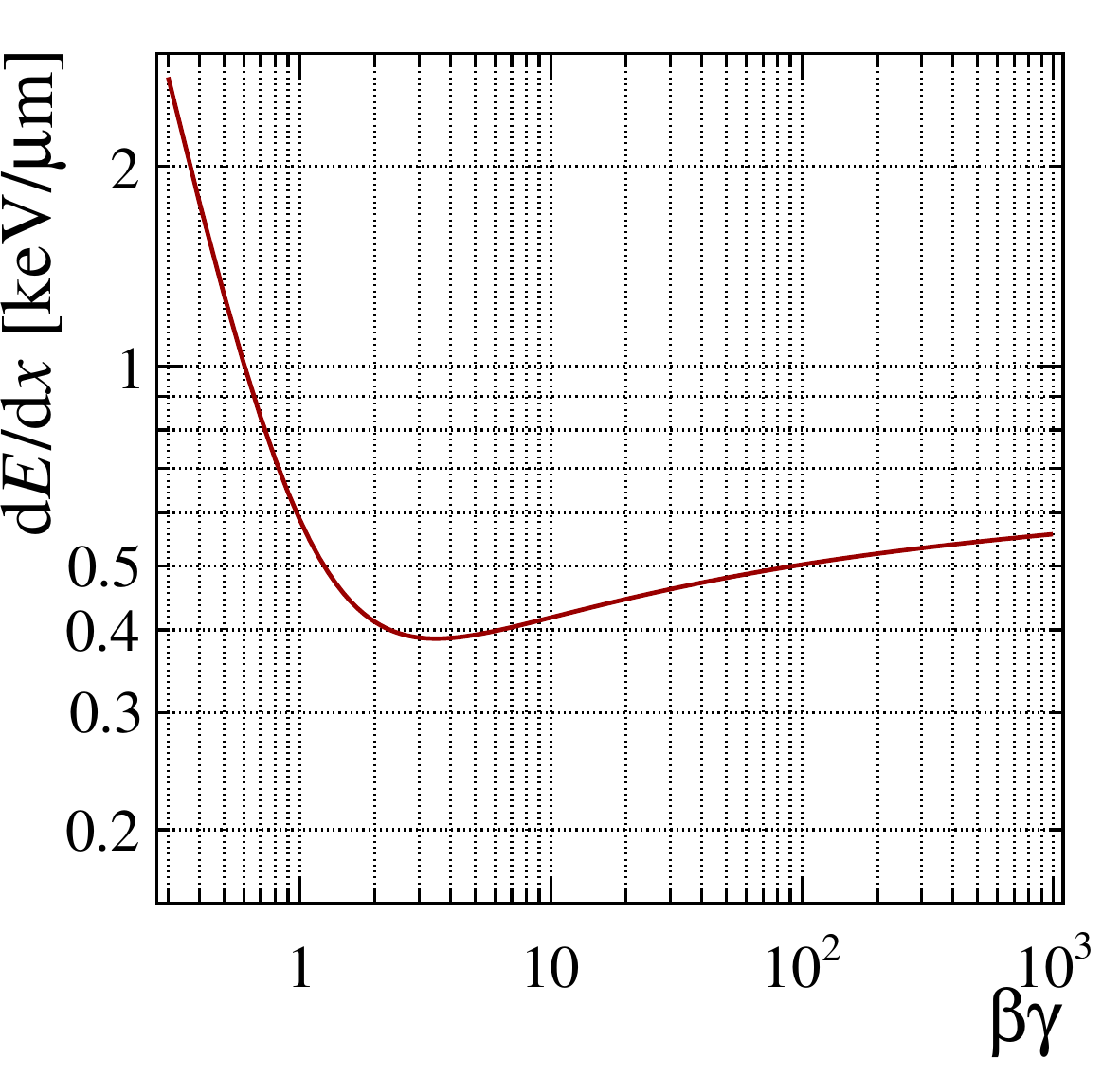}
   \caption{a) Average number of clusters per micrometer in silicon as a function of particle velocity.  b) Cluster size distribution for values of $\beta\gamma = 1$ (bottom), 4, 10, 100 (top) c) Average energy loss of a charged particle per unit of length in silicon. The numbers are  from Garfield++ \cite{garfield}  and Heed++ \cite{heed}.  }
  \label{charge_deposit}
  \end{center}
\end{figure}

\noindent
A charged particle interacts with the silicon sensor at discrete positions with an average distance of $\lambda$, which is shown in Fig. \ref{charge_deposit}a as a function of the particle velocity. 
For large particle velocities this number approaches a  constant value of $\lambda \approx 0.2\,\mu$m, equivalent to 5 clusters/$\mu$m. It has to be noted that this behaviour is not universal for all materials. As an example, for Argon gas the value of $\lambda$ increases again beyond a minimum at $\beta\gamma \approx 3$. The energy transferred in these interactions produces localized clusters of e-h pairs with a cluster size distribution $p_{clu}(n)$ shown in Fig. \ref{charge_deposit}b. We see that for large velocities the probability for larger clusters shows a slight increase. The resulting average energy loss per unit of length, typically approximated by the Bethe-Bloch function,  shows therefore a slight increase for large velocities, called  the 'relativistic rise' (Fig. \ref{charge_deposit}c).

\newpage


\section{General expression for the centroid time fluctuation in a silicon sensor}

\begin{figure}[ht]
 \begin{center}
     \includegraphics[width=7.5cm]{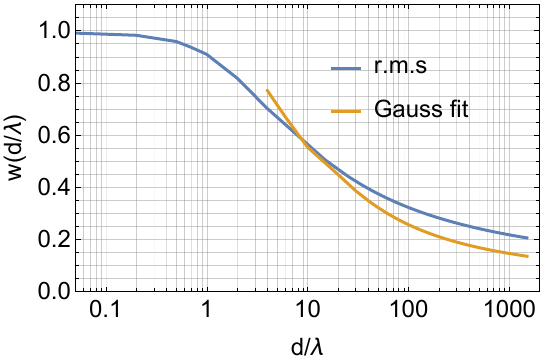}
   \caption{The function $w(d/\lambda)$ that determines the impact of charge deposit fluctuations (Landau fluctuations) on the time resolution of a silicon sensor. The particle velocity is  $\beta \gamma=10$. For thinner sensors, the impact of the Landau fluctuations is larger. Except for $d/\lambda \approx 10$ the distribution of the centroid times distributions are very non-Gaussian, so the r.m.s. differs significantly from the Gauss fit.   }
  \label{wplot}
  \end{center}
\end{figure}

\begin{figure}[ht]
 \begin{center}
     \includegraphics[width=7cm]{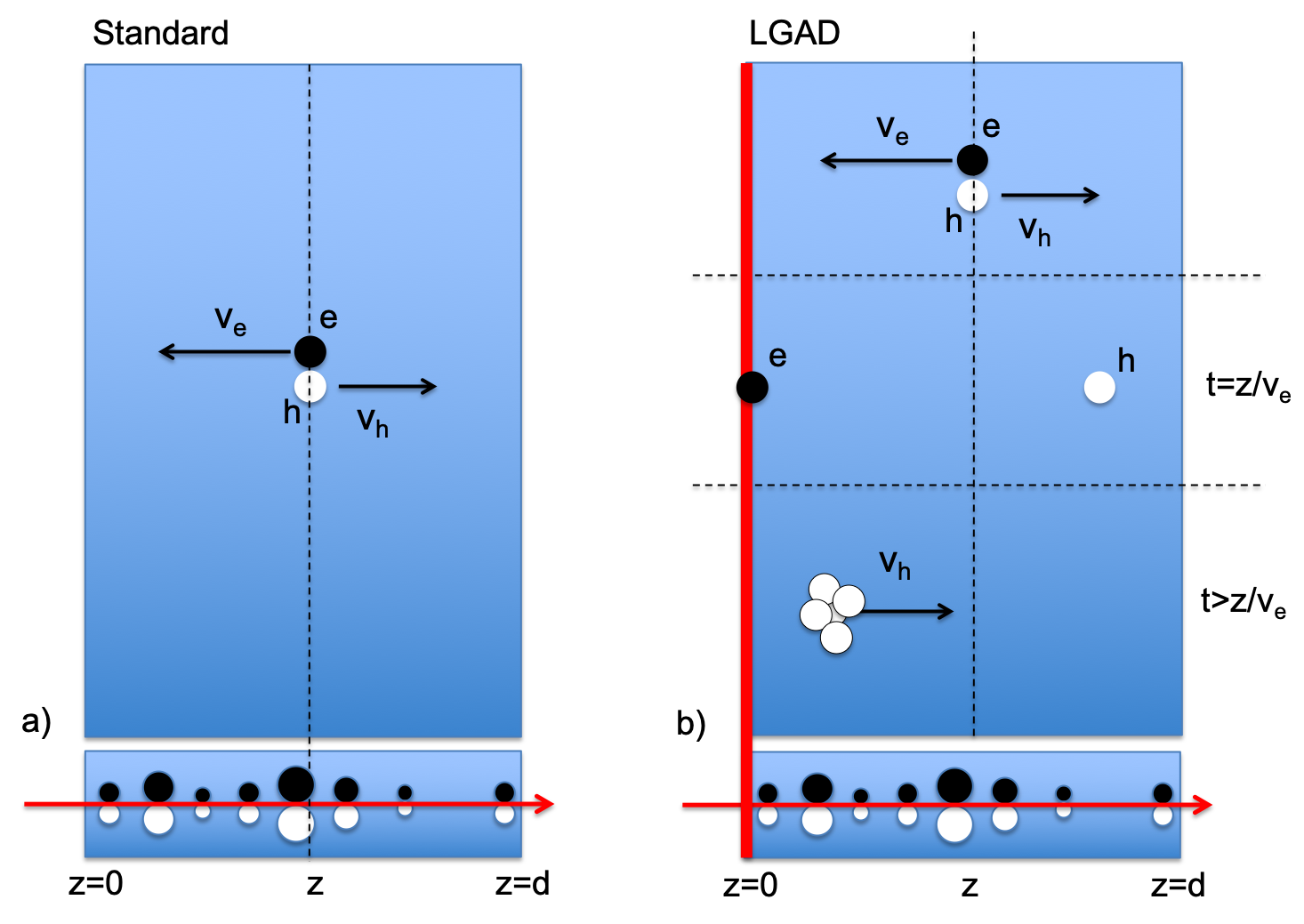}
   \caption{A single e-h pair and the ionization from a charged particle in a simplified geometry of a) a standard silicon sensor and b) a LGAD sensor. For the LGAD we assume an infinitely thin gain layer at $z=0$.   }
  \label{geometries}
  \end{center}
\end{figure}

\noindent
We assume a single e-h pair in a general sensor of thickness $d$ at position $z$ with $0{<}z{<}d$. Two specific examples with constant electric field are given in Fig. \ref{geometries}, but the following discussion holds for arbitrary  field dependence in the sensor. The movement of the electron and hole from this position $z$ will result in a signal $I(z, t)$. The centroid time of this signal is defined by
\beq \label{centroid_time}
   \tau_1 (z) = \frac{\int t I(z, t) dt}{\int  I(z, t) dt} 
\eeq
Assuming now a series of events, where single e-h pairs are randomly distributed between $z{=}0$ and $z{=}d$ with a uniform probability distribution, the average $\ov \tau_1$ and variance $\sigma_{\tau_1}$ of the centroid time are 
\beq \label{variance1}
    \ov \tau_1 = \frac{1}{d} \int_0^d \tau(z) dz \qquad  \ov{\tau_1^2} =  \frac{1}{d} \int_0^d \tau(z)^2 dz  \qquad \sigma^2_{\tau_1} =   \ov{\tau_1^2} -{\ov \tau_1}^2
\eeq
A charged particle crossing the sensor is not depositing a single e-h pair  but is creating clusters of e-h pairs at positions that are randomly distributed with a mean distance of $\lambda$. In Appendix 1 the following relation is shown: 
\\ \\
{\it If the variance of the centroid time for a uniform distribution of the position $z$ of a single e-h pair along $0{<}z{<}d$ is $\sigma_{\tau_1}$, the variance of the centroid time $\sigma_\tau$ due to the fluctuating charge deposit of a charged particle is
\beq \label{general_formula}
    \sigma_\tau   =  w \left( d/\lambda \right) \times \sigma_{\tau_1}
\eeq
where the function $w(d/\lambda)$ is a universal function that just depends on the cluster size distribution $p_{clu}(n)$ and the average distance between clusters $\lambda$.  }
\\ \\
The function $w(d/\lambda)$  is derived in  Appendix 1 (Eq. \ref{wlabel})  and it is represented in Fig. \ref{wplot}. 
The figure shows the same numbers as Fig. 4 of \cite{werner1}, but extended to $d{=}0$ and it is displayed as a function of $d/\lambda$. For relativistic particles  we have $\lambda{\approx} 0.2\,\mu$m, so for a silicon sensor of $50\,\mu$m thickness we have $d/\lambda \approx 250$. For a sensor thickness $d$ smaller than $\lambda$ there will always be a single cluster in the sensor,  $w(d/\lambda)$ is approaching unity and $\sigma_\tau = \sigma_{\tau_1}$. For larger sensor thickness the function $w(d/\lambda)$ is slowly decreasing. In \cite{werner1} it is shown that the Landau theory would result in an asymptotic behaviour of $w(d/\lambda) \approx 1/\sqrt{\ln d/\lambda}$. 
\\ \\
Fig. \ref{wplot} not only shows the r.m.s value from  Eq. \ref{wlabel} but also a Gaussian fit to a Monte Carlo evaluation of the centroid time distribution. 
The distributions of the centroid times are approximately Gaussian for values of $\lambda/d = 10$ but differ significantly from Gaussian distributions for other values. For values of $d/\lambda {>} 10$  the distributions have significant tails due to events with very large clusters. The standard deviation of a Gaussian fit is therefore smaller than the r.m.s. For values of $d/\lambda {<}10$ there are very few clusters in the sensor and the distribution approaches the one of a single cluster in the sensor. This distribution is highly non-Gaussian and it is calculated explicitly in the next section. Therefore we do not quote  Gaussian fit numbers for values of  $d/\lambda<4$. 
\\ \\
In conclusion, the function 
 $w(d/\lambda)$ increases slowly as the sensor thickness decreases, due to larger relative fluctuations in the charge deposit for thinner sensors. Therefore, the improved time resolution observed with thinner sensors is attributed to the reduction in $\sigma_{\tau_1}$, rather than a decrease in the fluctuations themselves. We will now apply these results to different sensor geometries.


\newpage

\section{Planar silicon sensors without gain}

\begin{figure}[ht]
 \begin{center}
     a)\includegraphics[width=7cm]{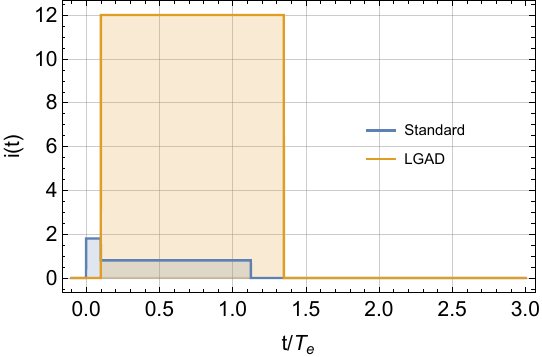}
     b)\includegraphics[width=7cm]{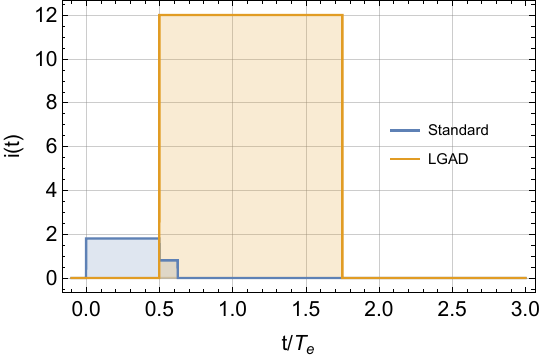}
      c)\includegraphics[width=7cm]{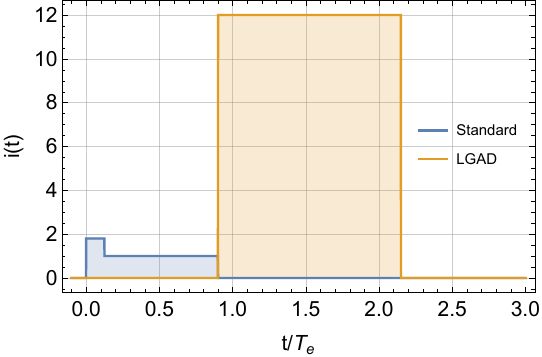}
  \caption{Silicon sensor signals due to a single e-h pair deposited at positions a) $z=0.1d$, b) $z=0.5d$, c) $z=0.9d$. We assume saturated drift-velocities with $v_h=0.8v_e$. 'Standard' refers to a standard planar sensor,  'LGAD' refers to the current from the holes produced in the avalanche of an LGAD with a gain of 15. For the LGAD, the signal is given by the sum of the two signals, and it is clearly dominated by the holes from the amplification.  }
  \label{shapes}
  \end{center}
\end{figure}

\begin{figure}[ht] \label{time_distributions} 
 \begin{center}
   a) \includegraphics[width=7cm]{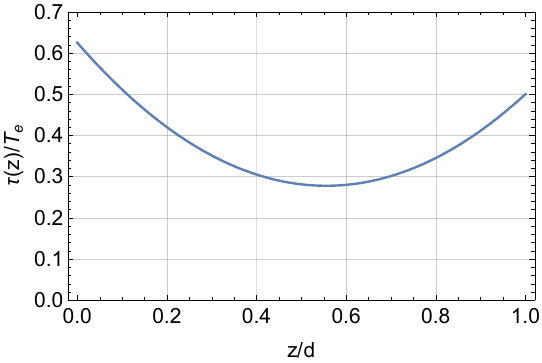} 
    b)\includegraphics[width=7cm]{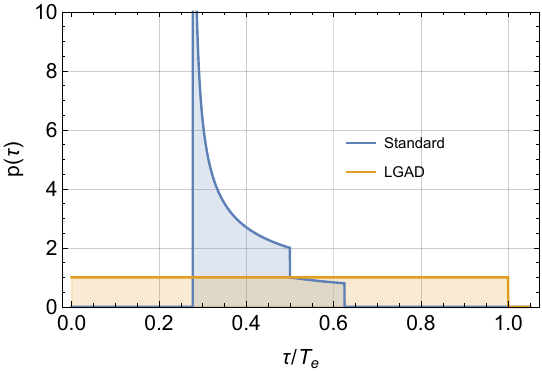}
   \caption{a) Signal centroid time versus position of the e-h pair. b) Centroid time distribution for a standard silicon sensor and an LGAD sensor, assuming a single e-h pair at a random position in the sensor as well as a drift-velocity ratio of $v_h=0.8v_e$. }
  \label{prob}
  \end{center}
\end{figure}

\noindent
We first assume a planar silicon sensor at very large over-depletion such that the electric field in the sensor can be assumed to be constant (Fig. \ref{geometries}a). The signal due to a single e-h pair deposited at position $z$ in the sensor is then given by 
\beq \label{standard_signal}
   I_S(z, t) = \frac{e_0 v_e}{d} 
   \Theta  \left(  z/v_e - t  \right)   
   +\frac{e_0 v_h}{d} 
   \Theta  \left[  (d-z)/v_h - t  \right]
\eeq
where $\Theta(t)$ is the Heaviside function. The electrons move in negative {z}-direction, the holes move in positve {z}-direction. We define $T_e=d/v_e$ and $T_h=d/v_h$, the times it takes electrons and holes to traverse the entire thickness $d$ of the silicon sensor. The signal is shown in Fig. \ref{shapes} for different positions of the primary e-h pair. The figure also shows the signal due to the movement of  the holes produced in the avalanche at $z=0$, assuming an infinitely thin gain layer and a gain of 15.
\\
The centroid time of this signal (Eq. \ref{centroid_time}) is then given  by 
\beq
      \tau_1(z) = \frac{1}{2d}\left[ \frac{z^2}{v_e} + \frac{(d-z)^2}{v_h}\right]
\eeq
Eq. \ref{variance1} then evaluates to 
\beq \label{sigma_single}
  \sigma_{\tau_1} = 
   \frac{1}{\sqrt{180}}\sqrt{
  4T_e^2 +4T_h^2    - 7T_eT_h
   }
\eeq
For this simple case we can give the explicit probability distribution for the centroid time $\tau_1$. The centroid time $\tau_1(z)$ is shown in Fig. \ref{prob}a, it has a minimum of $T=T_eT_h/(2(T_e+T_h))$ at $z=T_h/(T_e+T_h)\,d$ and assumes values of $\tau_1(d)=T_e/2$ and $\tau_1(0)=T_h/2$ at the boundaries. For a uniform distribution of the position $z$ we find with Eq. \ref{inversion} from Appendix 2 the distribution according to (Fig. \ref{prob}b).
\beq
      p(\tau_1) =\frac{1}{\sqrt{2(T_e+T_h)\tau_1 - T_e T_h}}
      \left[
     2 \Theta (\tau_1-T)
      - \Theta (\tau_1-T_e/2)
       - \Theta (\tau_1-T_h/2)
      \right]
\eeq
\\ \\
According to Eq. \ref{general_formula}, the variance of the centroid time for a charged particle is then 
\beq
   \sigma_{\tau} =   w(d/\lambda) \frac{1}{\sqrt{180}}\sqrt{
  4T_e^2 +4T_h^2    - 7T_eT_h
   }
\eeq
In case we have $T_e=T_h=T$, the time resolution is 
\beq  \label{standard_approx}
      \sigma_{\tau} =   w(d/\lambda) \frac{T}{\sqrt{180}}   \approx  0.075\,w(d/\lambda) \, T 
\eeq
These expressions were already derived in \cite{werner1}, however without relating it to the fluctuation of  single e-h pairs and only for situations where $d/\lambda \gg 1$. For a sensor of 50\,$\mu$m thickness at saturated drift-velocity of $v=10^7$\,cm/s this would evaluate to 7.5\,ps, which is the achievable time resolution in case of negligible noise jitter.


\section{Silicon sensors with gain, LGADs}

\noindent
In the so called Low Gain Avalanche Diode (LGAD) \cite{LGAD} a high field region is introduced in the sensor, where the electrons are multiplied.
We assume here an infinitely thin gain layer at $z=0$ and again a single e-h pair produced at position $z$. The movement of the electron and the hole from position $z$ produces the same signal as above, but when the electron arrives at $z=0$, it produces $G$ additional e-h pairs, and the holes are then moving through the entire sensor in positive {z}-direction. Therefore the total signal is equal to the sum of the two signals in Fig. \ref{shapes} and evaluates to 
\beq \label{LGAD_signal}
      I_L(z, t) =  I_S(z, t)
  + G \frac{e_0 v_h}{d} \left[
   \Theta(t-z/v_e)   - \Theta(t-z/v_e-d/v_h)
   \right]
\eeq
Eq. \ref{centroid_time} the evaluates to 
\beq
    \tau_1(z) = \frac{1}{G+1} 
    \left[
      \frac{1}{2d} \left(z^2/v_e+(d-z)^2/v_h\right) + G \times( z/v_e +d/2 v_h )
    \right]
\eeq
and Eq. \ref{variance1} evaluates to
\beq
     \sigma_{\tau_1} ^2 = \frac{1}{180(G+1)^2} 
   \left(
   4T_e^2+4T_h^2-7T_e T_h
   \right)
   +\frac{1}{12}
   \left(
    \frac{G}{G+1}T_e^2 - \frac{G}{(G+1)^2} T_eT_h
   \right)
\eeq
For a charged particle we have again $   \sigma_{\tau} = w(d/\lambda)  \sigma_{\tau_1} $. In the absence of gain i.e. $G \rightarrow 0$ we recover the expression from above for a standard planar sensor. For a gain in excess of around 10 the expression approximates to
\beq \label{LGAD_approx}
       \sigma_{\tau}\approx w(d/\lambda) \frac{T_e}{\sqrt {12}} \approx 0.3\,  w(d/\lambda) \, T_e
\eeq
This is the result when neglecting the signal from the primary e-h pair $I_S(z, t)$ in Eq. \ref{LGAD_signal} and the result is intuitive: in that case the LGAD signal is simply a 'box' of duration $T_h$ shifted by the time $t=z/v_e$, so a uniform distribution of $z$ in the interval $0<z<d$ results in a uniform distribution of $t$ in the interval $0<t<T_e$ (Fig. \ref{time_distributions}) and therefore an r.m.s time of $T_e/\sqrt{12}$. For a 50\,$\mu$m sensor at saturated drift-velocity of $v=10^7$\,cm/s this evaluates to 29\,ps, which is in the ballpark of the mesured values.
\\  \\
Comparing the approximate expressions from Eq. \ref{standard_approx} and Eq. \ref{LGAD_approx} we find that the 'Landau noise effect' of an LGAD is by a factor  $\sqrt{180/12}=\sqrt{15}\approx 3.9$ worse compared to a standard sensor. The big advantage of the LGAD is of course the fact that even for very thin sensors one has a large signal, providing an acceptable signal to noise ratio and therefore an 'elimination' of the noise jitter.


\section{3D sensors with planar electrodes}

\begin{figure}[ht] 
 \begin{center}
     \includegraphics[width=4cm]{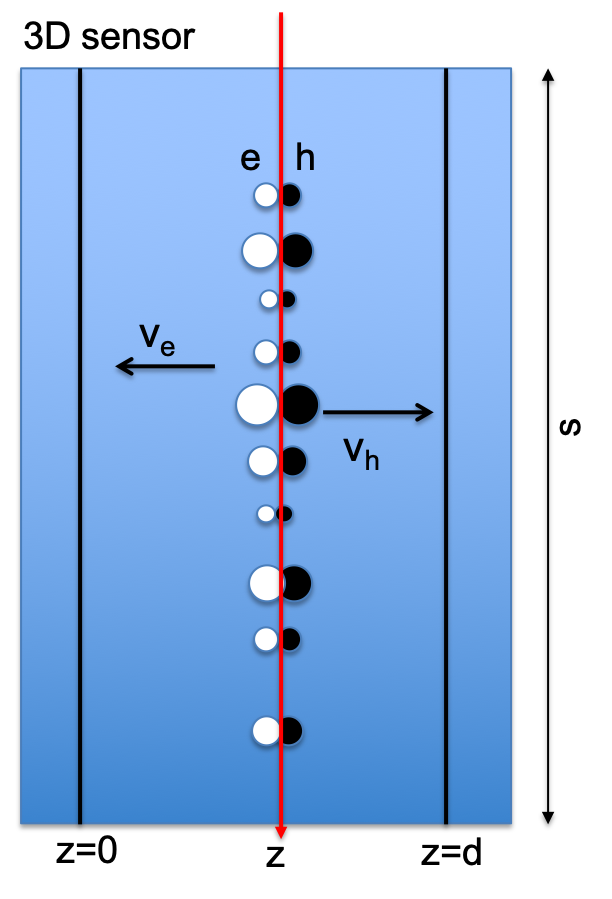}
   \caption{3D trench silicon detector. The time resolution is the one of a single e-h pair in a sensor of thickness $d$, while the charge deposit corresponds to a sensor of thickness $s$.  }
  \label{geometry_3D}
  \end{center}
\end{figure}

\noindent
In the so-called TIMESPOTproject \cite{timespot1, timespot2}, the readout electrodes are implemented as vertical trenches in the silicon sensor, realising a parallel plate geometry (Fig. \ref{geometry_3D}). The advantage is that one can realise a very thin sensor with an effective thickness of $d=10{-}20$\,$\mu$m and a charge deposit equivalent to $s{=}200\,\mu$m of silicon. If the particle is crossing the sensor perfectly parallel to the vertical trenches at some random position, the time resolution is simply equal to the number $\sigma_{\tau_1}$ from the single e-h pair in the sensor according to Eq. \ref{sigma_single}. The Landau fluctuations will only vary the pulseheight, which is removed by the normalization of the signal and therefore has no impact the timing. Inserting the saturated electron and hole velocities of $v_e=1.07\times 10^7$\,cm/s and $v_h=0.84\times 10^7$\,cm/s we find numbers of $\sigma_{\tau_1}=7.5/15$\,ps for an electrode distance of $d=10/20\,\mu$m. Despite the absence of Landau fluctuations, we find a finite time resolution due to the position dependence of the signal shape and therefore a fluctuation of the centroid time. 
It has to be noted that with $w(d/\lambda)=1$ for this case we find the  worst possible effect from the charge deposit fluctuations.
\\ \\
A slight tilt of the track will however produce Landau fluctuations with a very small effective average cluster distance $\lambda$ and therefore a very large number of $d/\lambda$ and in the following a small value of $w(d/\lambda)$. For example, assuming a sensor of $s=200\,\mu$m thickness with a trench distance of $d=20\,\mu$m and a slightly inclined track with an angle of  $\alpha= \arctan(d/s)$ we would effectively have a $d=20$\,$\mu$m sensor with an effective cluster distance of $\ov\lambda = \lambda/10$ of and therefore a $d/{\ov \lambda} = 1000$ and related $w(d/\ov \lambda) \approx 0.2$. 
The effect of Landau fluctuations on the time resolution should therefore have a significant dependence on the incidence angle of a particle. 


\section{Time resolution for short peaking times}

\noindent
In the previous sections have learned that the time resolution of a silicon sensor is equal to the fluctuation of the signal centroid time and independent of the threshold,  when using amplifier peaking times longer than the signal duration. We now investigate what happens for faster amplifiers with peaking times that are shorter than the signal duration, by first investigating the fluctuation of the signal shape due to Landau fluctuations. 
\\ \\
If we process the signal $I(z, t)$ from a single e-h pair at position $z$  with an amplifier of transfer function $h(t)$, the output signal is 
\beq
    G_1(z, t) = \int_0^t h(t-t') I(z, t') dt'
\eeq
The variance of the signal $\sigma_{G_1} (t)$ at time $t$ for a uniformly distributed random position $z$ of the e-h pair is then
\beq
           \ov G_1(t) = \frac{1}{d} \int_0^d G_1(z, t) dz \qquad     \ov {G_1(t)^2} = \frac{1}{d} \int_0^d G_1(z,t)^2 dz \qquad  \sigma_{G_1} (t)^2 =  \ov {G_1(t)^2} -   {\ov G_1(t)}^2
\eeq
As shown in Appendix 3, the variance of the signal for a charged particle is then 
\beq
        \sigma_{G}(t) = w(d/\lambda) \sigma_{G_1}(t)
\eeq
The argument follows the same line as the one for the fluctuation of the centroid time.
This fluctuation of the signal can now be pictured as 'Landau noise'  superimposed to the average signal, so if we apply a threshold to this signal we will find the associated time fluctuation by diving with the slope $\ov G'(t)$ of the average signal at the average threshold crossing time $t$
\beq
   \sigma_t(t) = \frac{\sigma_G(t)}{\ov G_1'(t)}
\eeq
As a final step we have to convert the time $t$ to the equivalent threshold $thr$ by using the average signal
\beq
       thr = \ov G_1(t) \qquad t = \ov G_1^{(-1)}(thr)
\eeq
and therefore find the time resolution as a function of the threshold. Applying these formulas for the  standard silicon sensor with the signal 
$I(z, t) = I_S(t, z)$ from Eq. \ref{standard_signal} and a preamplifier with delta response $h(t)$ and related transfer function $H(\om)$  \cite{blum_rolandi_riegler}
\beq \label{delta}
    h(t) = e^n n^{-n} \left(\frac{n\,t}{t_p} \right)^n \,e^{-nt\,/t_p} \qquad  H(\om) = \frac{t_p e^n n!}{(n+i\om t_p)^{n+1}}
\eeq
we find the time resolution as a function of threshold. The expressions for $G_1(z, t)$ is given in Appendix 4. Fig. \ref{leading_edge} shows $\sigma_t$ versus threshold for different amplifier peaking times with a choice of $n=2$. 
The time resolution is normalized to the centroid time resolution. As expected we find that for peaking times longer than the signal duration $T$, the time resolution is indeed approaching the centroid time resolution, and becomes independent of the threshold. For shorter peaking times there is a significant threshold dependence, and for low thresholds the time resolution can be better than the centroid time resolution, even tending to zero for zero threshold. It is clear that in absence of noise and with infinite bandwidth (i.e. $t_p \rightarrow 0$) one can achieve infinite time resolution due to the fact that induced current signal from Eq. \ref{standard_signal} is instantaneous and has and 'infinitely steep' leading edge at $t=0$. The question on the fastest applicable amplifier or the lowest possible threshold will be determined by the noise.
\begin{figure}[ht]
 \begin{center}
    \includegraphics[width=7cm]{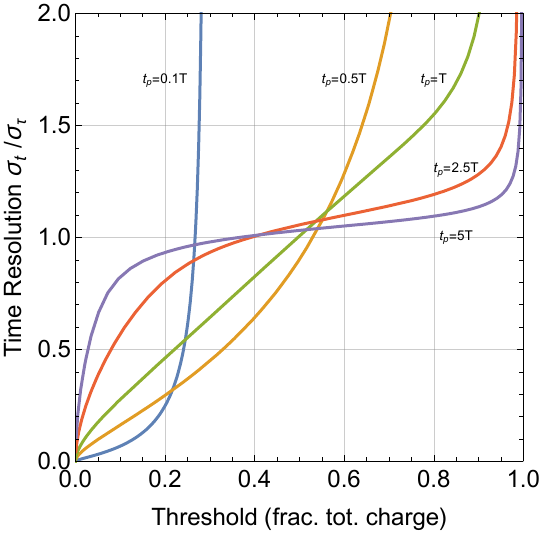}
   \caption{Time resolution as a function of  threshold for different amplifier peaking times for the standard silicon sensor, assuming $T_e=T_h=T$.  The time resolution is normalized to the centroid time resolution of $\sigma_\tau=w(d/\lambda)\times T/\sqrt{180} $. For long peaking times the time resolution becomes equal to the centroid time resolution and independent of threshold. For short peaking times there is a significant threshold dependence.}
  \label{leading_edge}
  \end{center}
\end{figure}
\\ \\
We can now perform the same procedure for the LGAD signal, but it is evident that for any choice of amplifier peaking time or threshold, the time resolution will always be equal to the centroid time resolution. When assuming only the gain signal from Eq.\ref{LGAD_signal}  the signal is simply a 'box' starting at time $t=z/v_e$ and having a duration of $T=d/v_h$. When convoluting this signal with any transfer function we will find a signal that does not vary in shape and is just randomly shifting by $t=z/v_e$, so at any applied threshold we will see the same time fluctuation as the one for the original current signal, which is equal to the centroid time fluctuation. In this approximation, the time resolution of an LGAD sensor cannot be improved beyond the centroid time resolution. The choice of amplifier peaking time $t_p$ will therefore mainly determined by the lowest achievable noise jitter.

\clearpage
\newpage


\section{Jitter from electronics noise, optimum filter}

\begin{figure}[ht]
 \begin{center}
    \includegraphics[width=7cm]{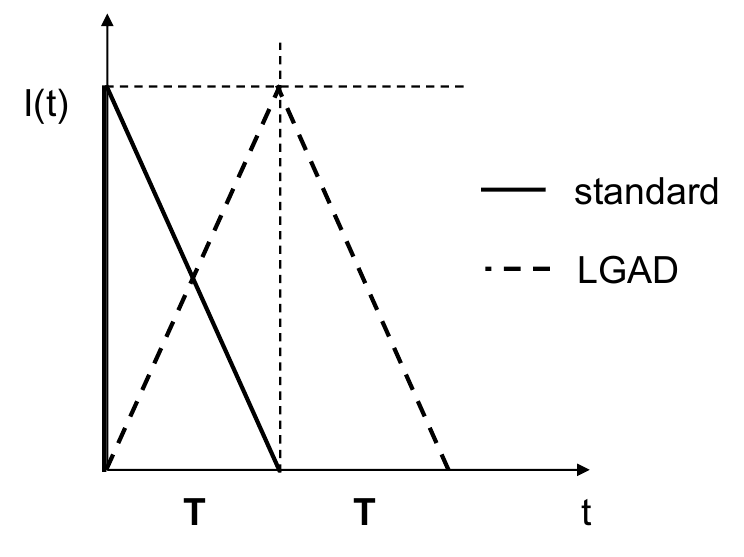}
   \caption{Average normalized signal of a standard silicon sensor and and LGAD sensor of the same thickness, assuming $T_e\approx T_h \approx T$. The average LGAD signal has twice the duration compared to the standard sensor signal.}
  \label{triangles}
  \end{center}
\end{figure}

\begin{figure}[ht]
 \begin{center}
     a)\includegraphics[width=7cm]{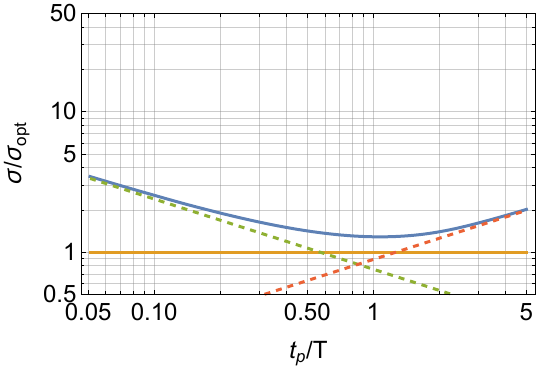}
     b)\includegraphics[width=7cm]{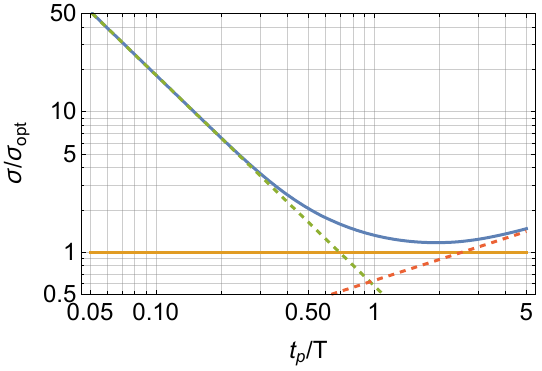}
   \caption{Timing jitter due to electronics noise for different amplifier peaking times, normalized to the lowest achievable value when using an optimum filter. The dashed lines show the approximations for large and small peaking time. a) Time jitter for the standard silicon sensor and b) time jitter for LGADs. We see that with an amplifier peaking time $t_p$ that is of the order of the total signal length, one can approach jitter values that are within 20-30\,\% of the lowest achievable ones with an optimum filter.}
  \label{real_filter}
  \end{center}
\end{figure}

\noindent
In addition to the Landau fluctuations there is of course the electronics noise that will contribute to the time resolution. Assuming a typical average signal shape from the silicon sensor and assuming only series noise related to the sensor capacitance $C$ we can use the theory of optimum filtering to minimise the time jitter. We first assume again a standard sensor at very high electric field, which is constant throughout the sensor and we assume equal electron and hole velocities $T_e\approx T_h \approx T$. The average signal has then triangular shape according to (Fig. \ref{triangles})
\beq
     i_S(t) = \frac{2Q_0}{T} \left(1-\frac{t}{T} \right) \qquad 0<t<T \qquad \int i_S(t)dt= Q_0
\eeq
where $Q_0$ is the total signal charge from primary ionization. The Fourier transform of the signal is
\beq
   I_S(\om) = \int_0^\infty i_S(t) \,e^{-i\om t} dt = \frac{2Q_0}{\om^2T^2} 
     \left(
     1-e^{-i\om T}-i\om T
     \right)
\eeq
Assuming a sensor capacitance $C$ and an amplifier with a white series noise power spectrum $e_n^2\,[V^2/Hz]$, the  noise power spectrum at the amplifier input is $w(\om)=e_n^2\om^2C^2$, and the maximum achievable slope to noise ratio is \cite{blum_rolandi_riegler}
\beq
   \left( \frac{k}{\sigma_V} \right)^2 = \frac{2}{\pi}\int_0^\infty \frac{\vert \om I_S(\om)\vert^2}{w(\om)}d\omega = \frac{Q_0^2}{e_n^2 C^2 } \,\frac{8}{3 T}
\eeq
which means that the minimum achievable time jitter is 
\beq
    \sigma_{S}^{opt} = \frac{\sigma_V}{k} = \frac{e_n C}{Q_0} \sqrt{\frac{3T}{8}}
\eeq
The average signal for the LGAD sensor in the approximation of  $T_e\approx T_h \approx T$ and operated a gain $G>10$ is (Fig. \ref{triangles})
\beq
     i_L(t) =\begin{cases}
     \frac{GQ_0}{T} \,\frac{t}{T}&  0<t<T \\
      \frac{GQ_0}{T} \left( 2-\frac{t}{T} \right) &  T<t<2T 
     \end{cases}
     \qquad \int i_L(t) dt = GQ_0 
\eeq
The Fourier Transform and minimum jitter are then 
\beq
     I_L(\om) = {\cal F} [i(t)] = \frac{2Q_0}{\om^2T^2} 
     e^{-i \om T}
     \left(
     1-\cos \om T
     \right)
     \qquad
         \sigma_{L}^{opt} =  \frac{e_n C}{GQ_0} \sqrt{\frac{3T}{4}}
\eeq
The factor $\sqrt{2}$ between the two expressions is due to the fact that the LGAD signal is twice longer than the signal from the standard sensor.
\\ \\ 
The required optimum filter to achieve this minimum time jitter is typically not realizable in practice, so we use a 'typical' transfer function of a charge sensitive amplifier according to Eq. \ref{delta} where we adapt the peaking time $t_p$ and filter order $n$ to give the best possible performance. The variance of the amplifier output noise for such a transfer function is then \cite{blum_rolandi_riegler}
\beq
      \sigma_V^2 (t_p)= \frac{1}{2\pi} \int_0^\infty w(\om) \vert H(\om)  \vert ^2 d \om=  e_n^2C^2  \frac{K_s^2}{t_p}  \qquad K_s^2=\frac{1}{2} \left( \frac{e}{2n}\right)^{2n}n^2 (2n-2)!
\eeq
The slope of the amplifier output signal $k(t)$ is given by 
\beq
    k(t) =  \int_0^t h'(t-t')i(t')dt'
\eeq
In order to minimize the time jitter we are setting the threshold to the point where the slope is maximal and we have a resulting timing jitter of 
\beq
      \sigma_j(t_p) =\frac{\sigma_V(t_p)}{k_{max}(t_p)} 
\eeq
We now vary the peaking time $t_p$ and  in order to minimize this number and we can compare to the minimum jitter that an optimum filter would provide. We don't give the analytic expressions but just show the results in Fig. \ref{real_filter}. We find the for amplifier peaking times that are similar to the signal duration $T$ we can achieve jitter values that are within 20-30\,\% of the minimum achievable ones. For peaking times of $0.5T<t_p<5T$ the jitter is within a factor of two of the lowest achievable one. For very long and very short peaking times we can find explicit expressions as shown in the following.
\\ \\
For amplifier peaking times much longer than the signal time i.e. $t_p \gg 2T$ the output signal is simply the delta response multiplied by the charge of the signal. The maximum slope of the delta response in Eq. \ref{delta}  is 
\beq
    k_{S} = Q_0  \frac{K_p}{t_p}  \qquad  k_{L} = G Q_0  \frac{K_p}{t_p}  \qquad K_p =e^{\sqrt{n}}n^{3/2-n}(n-\sqrt{n})^{n-1} 
 \eeq
The time jitter  $\sigma= \sigma_V/k$, related it to the optimum jitter, is then 
\beq
  \frac{\sigma_S}{\sigma_S^{opt}} = \frac{K_s}{K_p} \sqrt{\frac{8}{3}} \sqrt{t_p/T} \qquad   
  \frac{\sigma_L}{\sigma_L^{opt}} = \frac{K_s}{K_p} \sqrt{\frac{4}{3}} \sqrt{t_p/T}  
 \eeq
For both, standard silicon sensor and LGADs, the  jitter increases with $\sqrt{t_p}$ and these expressions are represented in Fig. \ref{real_filter}.
\\ \\
For very short peaking times $t_p \ll T$, the expressions for the maximum slope approximate to 
\beq
     k_{S} = \frac{2Q_0}{T}  \qquad  k_{L}  =G Q_0(e^n n^{(-1 - n)}n! ) \frac{ t_p}{T^2} 
\eeq
and we therefore have 
\beq
       \frac{\sigma_{S}}{\sigma_S^{opt}} =K_s \sqrt{\frac{2}{3}} \frac{1}{\sqrt{t_p/T}} \qquad  
       \frac{\sigma_{L}}{\sigma_L^{opt}} =K_s \sqrt{\frac{4}{3}} \frac{1}{e^n n! n^{-n-1}} \,\frac{1}{(t_p/T)^{3/2}} 
\eeq
For the standard silicon sensor the jitter value varies as $\propto 1/\sqrt{t_p}$ while for the LGAD sensor the jitter shows a much stronger dependence  $\propto 1/(t_p)^{3/2}$. This is clearly due to the fact that for a standard silicon sensor the largest current is at $t=0$ so also for short peaking times one integrates significant charge. For the LGAD sensor the current at $t=0$ is equal to zero, so the slope to noise ratio for small peaking times becomes small and the jitter diverges quickly.
\\ \\
We can conclude that for minimizing  the time jitter due to series noise we have to use amplifiers with a peaking time that are similar to the signal duration. For the LGAD sensor the jitter has a strong increase for shorter peaking times.

\clearpage
\newpage


\section{Correlation between of the effects of noise and Landau fluctuations}

\noindent
Since the electronics noise and the Landau fluctuations are not correlated one would now be tempted to perform the square sum of the two contributions to find the final performance number. However, the combined effect of noise and Landau fluctuations will typically provide a worse performance number than the simple square sum, as illustrated by the following example. \\
We assume a simplified situation, where a signal shows a linearly rising leading edge according to $I(t) = k \times t$ and we apply a threshold $v$ to the signal. The measured threshold crossing time is then $t=v/k = r \times v$, where we have defined the inverse slope $r=1/k$. The slope will now vary from event to event and we can formulate the electronics noise  as a variation of the threshold $v$ from event to event. We therefore assume the slope (or inverse slope) and the threshold to vary according to some probability distribution with a mean and variance of $\ov r, \ov v$ and $\sigma_r, \sigma_v$. If we assume only the effect of the slope variation at fixed threshold $\ov v$ we find a time resolution of $\sigma_t =  \ov v \sigma_r$. If we consider only the noise, i.e. if we assume a varying threshold with the average slope $\ov r$ we find a time resolution of $\sigma_t= \ov r \sigma_v$, so the square sum is
\beq
      \sigma_t^2 =   \ov v^2 \sigma_r^2 + \ov r^2 \sigma_v^2
\eeq
The correct variance of the threshold crossing time $t= v \times r$ however is
\beq
    \sigma_t^2 =  \ov v^2 \sigma_r^2 + \ov r^2 \sigma_v^2 + \sigma_r^2  \sigma_v^2 
\eeq
so it is larger than the square sum of the individual effects by an amount of $\sigma_r^2  \sigma_v^2$. As illustrated by this simple example we can in general not expect that the time resolution of a sensor can be decomposed into the square sum of individual contributions resulting from  Landau fluctuations and noise. 


\section{Conclusion}

\noindent
We have given explicit expressions for the impact of Landau fluctuations and electronics noise on the time resolution of standard, LGAD and 3D  trench silicon sensors. 
\\ \\
The effect of Landau fluctuations was related to the time fluctuations for a single e-h pair through a general function $w(d/\lambda)$ that is calculated in an exact way. For values of $0<d/\lambda<1000$ the function $w$ varies from between 1 and 0.2. The strong improvement of the time resolution for thin sensors is due to the reduced fluctuations of the centroid time from the decreased total electron and hole drift times. In the approximation of equal electron and hole velocities and assuming amplifier peaking times longer than the signal length, the expressions are
\bea
      \mbox{Standard: } \sigma_\tau &= & w(d/\lambda) \frac{T}{\sqrt{180}}    \\
          \mbox{LGAD: }  \sigma_\tau &=&  w(d/\lambda) \frac{T}{\sqrt{12}}    \\
        \mbox{Trench: }     \sigma_\tau &=   &\frac{T}{\sqrt{180}}   
\eea
A comparison of the resulting expressions with measurements was already given in \cite{werner1}. 
\\ \\
In absence of noise, the time resolution for standard and trench sensors can be arbitrarily improved by faster electronics and lower thresholds. For LGAD sensors, when neglecting the signal from the primary charges,  the time resolution will always be equal to the centroid time resolution, independent of the amplifier bandwidth and threshold.
\\ \\
When considering the noise jitter only we find the minimum values for amplifier peaking times that are similar to the duration of the total signal length. In this case one can achieve jitter values that are within 20-30\% of the ones achievable with optimum filters. This will represent the optimum for the LGAD sensors.
For standard and trench sensors the optimum peaking time will be a compromise between the increase of noise jitter and decrease of Landau noise when reducing the peaking time.
\\ \\
The calculations presented in the report give the principal mechanisms that limit the time resolution of silicon sensors and the simplified geometries can be used as benchmarks for more detailed simulations. The energy deposit, drift and multiplication of charges in silicon sensors can be very efficiently be simulated with programs like garfield++ \cite{garfield}. For a given sensor configuration one can  calculate a sample of induced current signals and then perform a sweep of amplifier parameters, noise levels and thresholds to find out the optimum processing parameters.


\section{Acknowledgments}

\noindent
We would like to thank Heinrich Schindler for providing the data of the PAI model as well as Giuseppe Iacobucci and his group from University of Geneva for many interesting discussions, specifically Matteo Milanesio for many important suggestions.

\newpage


\section{Appendix 1} 

\noindent
A charged particle passing a sensor of thickness $d$ interacts with the material at discrete positions with a mean distance of $\lambda$, and it creates clusters of e-h pairs at these positions according to a cluster size distribution $p_{clu}(n)$. The total number of clusters therefore follows a Poisson distribution with a mean of $d/\lambda$. Under the assumption of a fixed number of clusters in the sensor, the position of the clusters follows a uniform distribution, so if we calculate the variance of the centroid time $\sigma_{\tau_m}^2$ for $m$ uniformly distributed clusters in the sensor, we can find the variance of the centroid time for a Poisson distribution of clusters $P(m)$ by $\sigma_\tau^2 = \sum P(m) \sigma_{\tau_m}^2$. This calculation is performed in the following.
\\ \\
We assume $I_1(z, t)$ to be the sensor signal due to a single electron-hole pair deposited at position $z$ inside the sensor. The 'centroid time' is then defined by
\beq
      \tau_1(z) = \frac{\int t I_1(z, t) dt}{\int I_1(z, t) dt}
\eeq
The average and variance of the centroid time for a uniform distribution of the e-h pairs in the sensor along $0{<}z{<}d$ is then
\beq 
    \ov \tau_1 = \frac{1}{d} \int_0^d \tau_1(z) dz \qquad  \ov{\tau_1^2} =  \frac{1}{d} \int_0^d \tau_1(z)^2 dz  \qquad \sigma^2_{\tau_1} =   \ov{\tau_1^2} -{\ov \tau_1}^2
\eeq
Assuming two charge deposits in the sensor at position $z_1$ and $z_2$ with charges $n_1$ and $n_2$, the centroid time is 
\beq
       \tau_2(z_1, z_2) = \frac{n_1 \tau_1 (z_1) + n_2 \tau_1(z_2)}{n_1+n_2}
\eeq
If the positions $z_1$ and $z_2$ are uniformly distributed and the cluster size distribution is $p_{clu}(n)$, the average of the centroid time is 
\beq
     \ov \tau_2 =\sum_{n_1=1}^\infty \sum_{n_2=1}^\infty p_{c}(n_1)p_{c}(n_2)   \frac{1}{d^2}\int_0^d \int_0^d\tau_2(z_1,z_2) dz_1dz_2 = \ov \tau_1 
\eeq
and the second moment is
\bea
     \ov{ \tau_2^2} & = & \sum_{n_1=1}^\infty \sum_{n_2=1}^\infty p_{c}(n_1)p_{c}(n_2) \frac{1}{d^2}\int_0^d \tau_2(z_1,z_2) ^2dz_1dz_2 \\
                                          & = &              ( {\ov \tau_1}^2 -\ov \tau_1^2 )   \sum_{n_1=1}^\infty \sum_{n_2=1}^\infty p_{c}(n_1)p_{c}(n_2)   \frac{n_1^2+n_2^2}{(n_1+n_2)^2}    +\ov \tau_1^2 
\eea
The variance therefore is
\beq
    \sigma_{\tau_2}^2  
       =   \sigma_{\tau_1}^2 \sum_{n_1=1}^\infty  \sum_{n_2=1}^\infty  \frac{2 n_1^2 p_{c}(n_1)}{(n_1+n_2)^2}p_{c}(n_2) 
\eeq
In general for $m$ primary clusters we have 
\beq
      \sigma_{\tau_m}^2   = \sigma_{\tau_1}^2   \sum_{n=1}^\infty  \sum_{n_1=1}^\infty   \frac{m\, n_1^2\, p_{c}(n_1)}{(n_1+n)^2} p(n, m-1) 
\eeq
where we have defined $p(n, m)$ as the probability to find a total number of $n$ e-h pairs in the sensor in case there are $m$ primary clusters in the sensor.
The probability to find $m$ primary clusters in the sensor, considering only efficient events with $m>0$, follows a Poisson distribution, so we have   
\beq
      P(m) = \frac{1}{e^{d/\lambda}-1}\frac{   \left( \frac{d}{\lambda} \right)^m }{m!}
\eeq
Because it holds that the average is the same for each number of clusters, i.e. $\ov \tau_m = \ov\tau_1$ we can write the full variance for the centroid time as 
\bea
      \sigma_{\tau}^2 & =  &  \sum_{m=1}^{\infty} P(m)   \sigma_{\tau_m}^2   \\
        & =  &  \sigma_{\tau_1}^2 \left[ P(1)+\sum_{m=2}^{\infty} P(m)  \left(\sum_{n=1}^\infty  \sum_{n_1=1}^\infty   \frac{m\, n_1^2\, p_{c}(n_1)}{(n_1+n)^2} p(n, m-1) \right)\right] \\
                    & =  &  \sigma_{\tau_1}^2\frac{d}{\lambda} \left[ \frac{ 1}{e^{d/\lambda}-1}+\sum_{n=1}^\infty  \sum_{n_1=1}^\infty   \frac{\, n_1^2\, p_{c}(n_1)}{(n_1+n)^2}  p(n,d) \right] \\
                          & = &  \sigma_{\tau_1}^2 w \left( d/\lambda  \right) ^2\\
\eea
where we have $p(n, d)$ to be the probability to find a total number of n e-h pairs in the sensor of thickness $d$ and we have defined 
\beq \label{wlabel}
     w \left( d/\lambda  \right) ^2 = \frac{d}{\lambda} \left[ \frac{ 1}{e^{d/\lambda}-1}+\sum_{n=1}^\infty  \sum_{n_1=1}^\infty   \frac{\, n_1^2\, p_{c}(n_1)}{(n_1+n)^2}  p(n,d) \right] 
\eeq
In case we have $d \gg \lambda$ we can replace the sums by integrals and the expression approximates to
\beq
      w \left( d/\lambda \right)^2  \approx \frac{d}{\lambda} \int_{0}^\infty \left[  \int_{0}^\infty   \frac{\, n_1^2\, p_{c}(n_1)}{(n_1+n)^2}   dn_1  \right] p(n,d)  dn
\eeq
which is equal to the expression from \cite{werner1}.


\section{Appendix 2}

\noindent
Assuming a monotonic function $\tau=f(z)$ with a random distribution of $z$ according to a probability distribution $p_z(z)$, the resulting probability distribution $p_\tau(\tau)$ for $\tau$ is given by
\beq \label{inversion}
    p_\tau(\tau) = \frac{1}{f'(f^{-1}(\tau))} p_z(f^{-1}(\tau))
\eeq
In case the function $f(z)$ is not monotonic, the different branches of the function have to be inverted separately and the added together.


\section{Appendix 3}

\noindent
The fluctuation of the full signal shape of a detector signal can be calculated the following way. We assume $G_1(z, t)$ to be the signal due to a single e-h pair produced at position $z$ in the sensor as being processed by the readout electronics, i.e.
\beq
     G_1(z, t) = \int_0^t f(t-t')I_1(z, t')dt'
\eeq 
with $I_1(z, t)$ being the induced current from this single e-h pair. The average and the variance of the signal at time $t$ for a random uniform distribution of the position $z$ is then
\beq
           \ov G_1(t) = \frac{1}{d} \int_0^d G_1(z, t) dz \qquad     \ov {G_1(t)^2} = \frac{1}{d} \int_0^d G_1(z,t)^2 dz \qquad  \sigma_{G_1} (t)^2 =  \ov {G_1(t)^2} -   {\ov G_1(t)}^2
\eeq
Assuming two charge deposits in the sensor at position $z_1$ and $z_2$ with charges $n_1$ and $n_2$, the signal (normalized to the total charge) is 
\beq
       G_2(z_1, z_2, t) = \frac{n_1 G_1 (z_1, t) + n_2 G_1(z_2, t)}{n_1+n_2}
\eeq
From here the argument proceeds the same way as for the centroid time and we find an r.m.s. fluctuation of the normalized signal for a charged particle as
\beq
        \sigma_{G}(t) = w(d/\lambda) \sigma_{G_1}(t)
\eeq
%
%
%


\section{Appendix 4}

\bea
    \frac{d}{t_p}\frac{n^{n+1}}{e^n}\,G_1(z,t) 
   & = &
     v_1\, \Theta(z-v_1t) \, [n!-\Gamma(n+1,t/t_p)]  \no  \\ 
   &-& 
    v_1 \,\Theta(v_1t-z) \, [\Gamma(n+1,t/t_p)-\Gamma(n+1,-(z-v_1t)/(t_pv_1)] \no \\
   &+&
   v_2\, \Theta((d-z)-v_2t) \, [n!-\Gamma(n+1,t/t_p] \no \\
   &-& 
   v_2 \,\Theta(v_2t-(d-z)) \, [\Gamma(n+1,t/t_p)-\Gamma(n+1,-(d-z-v_2t)/(t_pv_2)]  
   \no    
\eea


\end{document}